\documentclass[aps,preprint,floatfix,showpacs,superscriptaddress]{revtex4}
\usepackage{graphicx}
\usepackage{amsmath}
\usepackage{bm}

\newcommand{\be}{\begin{equation}}
\newcommand{\ee}{\end{equation}}
\newcommand{\bel}[1]{\be\label{#1}}
\newcommand{\re}[1]{Eq.~(\ref{#1})}

\newcommand{\ds}{\displaystyle}
\newcommand{\ov}[1]{\overline{#1}}
\newcommand{\hsp}{\hspace*{1pt}}

\begin{document}

\title
{Possible production of exotic baryonia\\
in relativistic heavy--ion collisions}


\author{I.N. Mishustin}

\affiliation{The Kurchatov Institute, Russian Research Center,
123182 Moscow, Russia}

\affiliation{Institut~f\"{u}r Theoretische Physik,
J.W. Goethe Universit\"{a}t,\\
D--60054 Frankfurt am Main, Germany}

\affiliation{The Niels Bohr Institute, University of Copenhagen, DK--2100 
Copenhagen {\O}, Denmark}

\author{L.M. Satarov}

\affiliation{The Kurchatov Institute, Russian Research Center,
123182 Moscow, Russia}

\affiliation{Institut~f\"{u}r Theoretische Physik,
J.W. Goethe Universit\"{a}t,\\
D--60054 Frankfurt am Main, Germany}

\author{D. Strottman}

\affiliation{Institut~f\"{u}r Theoretische Physik,
J.W. Goethe Universit\"{a}t,\\
D--60054 Frankfurt am Main, Germany}

\affiliation{Los Alamos National Laboratory,
87545 New Mexico, USA}

\author{W. Greiner}

\affiliation{Institut~f\"{u}r Theoretische Physik,
J.W. Goethe Universit\"{a}t,\\
D--60054 Frankfurt am Main, Germany}

\affiliation{Frankfurt Institute for Advanced Studies,
J.W. Goethe Universit\"{a}t,\\
D--60054 Frankfurt am Main, Germany}

\begin{abstract}
Properties of a hypothetical baryonium with the quark
content ($uds\ov{u}\ov{d}\ov{s}$) are discussed. The MIT bag
model predicts its mass to be unexpectedly low, approximately
1210 MeV. Possible hadronic decay modes of this state are analyzed.
Ultrarelativistic heavy--ion collisions
provide favorable conditions for the formation of such particles from
the baryon--free quark--gluon plasma. We estimate multiplicities of
such exotic baryonia on the basis of a simple thermal model.
\end{abstract}


\maketitle

Recent experimental observations of exotic baryons, namely the $\Theta^+
(1540)$~\cite{Nak03} and $\Xi^{--} (1862)$~\cite{Alt04} pentaquarks,
are of great interest for understanding strong interactions. On the
other hand, many predictions of exotic hadrons, made within different
effective models of QCD are still waiting for their experimental
verification.  For instance, the dibaryon state with the quark content
($u^2d^2s^2$) has been predicted~\cite{Jaf77a} using the MIT bag
model~\cite{Cho74} and not found yet. In
Refs.~\cite{Jaf77b,Str78,Str79} this model was used to calculate masses
of exotic multi-quark-antiquark states, in particular, the pentaquarks
($q^4\ov{q}$) and heptaquarks~($q^5\ov{q}^2$). Later, low-lying
pentaquark states have been predicted within the chiral soliton
model~\cite{Dia97}.

In Ref.~\cite{Mis00} binding energies of multi-$q\ov{q}$ systems with
different flavor compositions have been studied within a completely
different approach. Namely, the equation of state of $q\ov{q}$ matter
at fixed relative concentrations of antiquarks was calculated within
the generalized Nambu--Jona-Lasinio model. It was shown that such
systems have bound states at zero pressure. The maximum binding energy
per particle was obtained for systems with equal number of quarks and
antiquarks, so-called ''mesoballs''. In these systems the repulsive
vector interaction vanishes (at least on the mean--field level) and the
binding energy is determined by a competition between the scalar
attraction and the kinetic energy of quarks and antiquarks. The kinetic
(Fermi) energy is lowered by transforming a fraction of light $q\ov{q}$
pairs into strange $s\ov{s}$ pairs. Such flavour symmetrization is also
favored by the instanton--induced flavor-mixing interaction. As a
consequence, the binding energy per quark--antiquark pair is predicted
to be maximal for systems with large hidden strangeness. In these
mean--field calculations the flavor and color correlations are
considered only on average, and therefore, clusterization of
$q\ov{q}$ matter is ignored. Possible formation of
multi-quark-antiquark systems was suggested also in Ref.~\cite{Mis04}
in connection with deeply bound antibaryon states in nuclei.
The role of attractive color
configurations of quarks and gluons and possible existence of their
bound states in the quark--gluon plasma have been discussed in
Refs.~\cite{Shu04}.

Recently one of us reconsidered~\cite{Str04} the systematics of
multi--$q\ov{q}$ clusters by using the MIT bag model with the same set
of parameters as in Ref.~\cite{Cho74}. In particular, masses of exotic
systems with the quark content \mbox{$(u\ov{u})^m (d\ov{d})^n
(s\ov{s})^k$}, where \mbox{$m+n+k=3,4,5,6$}, have been calculated.  We
use the name exotic baryonia for states with \mbox{$m+n+k=3$} in order
to distinguish them from the $B\ov{B}$ bound states discussed long ago
within the mesonic picture of strong interaction (see
Refs.~\cite{Sha78,Ric00}). Recent measurements~\cite{Bai03} of the
$p\ov{p}$ mass spectrum in $J/\psi\to\gamma p\ov{p}$ decays seem to
give experimental evidence in favor of the subthreshold $N\ov{N}$ bound
state with mass $m\simeq 1859$\,MeV. This observation has renewed
interest in the issue of baryonia. For instance, in Ref.~\cite{Dat03}
the $\Lambda\ov{\Lambda}$ baryonium with mass $m\simeq {2200}$ MeV has
been proposed.

In our opinion, the above mentioned loosely bound states are not the
lowest energy states of baryonia. According to the MIT bag calculations
of Refs.~\cite{Str04,Buc79} the lightest $q^3\ov{q}^{\hsp 3}$ state is a
flavor symmetric pseudoscalar ($J^P=0^-$), with the quark content
($u\ov{u}\hsp d\ov{d}\hsp s\ov{s}$). Such state has zero isospin, zero
strangeness and positive G-parity. For brevity we call it $Y$-particle.
Its mass is predicted to be unexpectedly small, \mbox{$m_Y\simeq
1214$\,MeV}, i.e. significantly below the $\Lambda\ov{\Lambda}$
threshold (2230 MeV). It has such a low mass because of the
unique arrangement of quarks and antiquarks leading to a very strong
color-magnetic interaction~\footnote{ The calculated radius of the 
$Y$-particle is rather small, about 0.87~fm, that corresponds to the 
density of quarks and antiquarks of about 2.2~fm$^{-3}$. 
}.
Within the Jaffe approximation~\cite{Jaf77b}, the
mass-dependent factors in the interaction terms are replaced by
average values. Then the color-spin interaction reduces to terms that
can be expressed via Casimir operators of $SU(6)_{cs}, SU(3)_c$\,, and
$SU(2)_s$\,. To obtain antisymmetric wave functions, the
flavor $SU(3)_f$ symmetry representation should be conjugate to the
$SU(6)_{cs}$ representation.  The larger the symmetry of the $3q$ or
$3\ov{q}$ wave functions in $SU(6)_{cs}$\,, the greater is the
color-magnetic attraction. The three-quark states that are completely
symmetric (i.e., the [3]-states) do not contain $SU(3)_c$
singlets~\cite{Str79} and hence, they can not exist as free particles.
Only by coupling to three antiquarks with the same symmetry, one can
produce a color singlet state. Thus, the ($uds$) and
($\ov{u}\ov{d}\ov{s}$) combinations in the $Y$-particle are not color
singlets, unlike in the quasimolecular $\Lambda\ov{\Lambda}$ state.

Let us discuss possible decay modes of the $Y$-particle.  First of all
we note that such simple channels as \mbox{$Y\to K\ov{K}$} and
\mbox{$Y\to n\pi$} with an even number of secondary pions are forbidden
due to the parity and angular momentum conservation. The decays into
odd numbers of pions should be suppressed because of the G-parity
violation. Thus, we expect that only the 3-body decay channels
\mbox{$Y\to\pi K\ov{K}$} and  $Y\to 2\pi\eta$ may contribute
significantly to the width. These final states can be obtained by
simple rearrangement of quarks from the initial state.
However, it is difficult to calculate the corresponding
matrix elements. If they are of the same order, the relative
contribution of these two channels is controlled by the phase space
available for final particles. The relativistic
phase space volume corresponding to the decay of a
particle with mass $m$ into $n$ particles with
masses~$m_1,\ldots,m_n$\,is defined as
\bel{phsv}
R_n=\int\prod\limits_{i=1}^{n}\frac{d^3k_i}
{k_i^0}\hsp\delta^{(4)}(P-\sum\limits_{l=1}^{n}k_l)\,,
\ee
where $P^\mu=(m,\bm{0})^{\mu}$ and
$k_i^\mu=(\sqrt{m_i^2+\bm{k}_i^2},\bm{k}_i)^\mu$ are, respectively, the
4-momenta of initial and final particles (in the rest frame of the
decaying particle). The direct calculation gives
\bel{deck} \frac{\ds
R_3(Y\to\pi K\ov{K})}{R_3(Y\to 2\pi\eta)} \simeq 0.0543\,.
\ee
One can see that the $Y\to\pi K\ov{K}$ decay channel is strongly
suppressed due to closeness to the threshold. On the other hand, the
$\eta$ meson has smaller hidden strangeness than the $K\ov{K}$ pair,
and thus, the $Y\to 2\pi\eta$ decay mode may be additionally suppressed
by the OZI rule. From the experimental viewpoint, using this decay
channel is problematic because of the small width of $\eta$ meson.
We would like also to point out that, similarly to $\eta$ meson, the
$Y$-particle should have a significant $\gamma\gamma$ width, most
likely, in the keV range. Therefore, the observation of exotic
$Y$-baryonium in $pp$ or $\pi p$ reactions should be possible
by measuring the $\pi K\ov{K}$, $2\pi\eta$ and $\gamma\gamma$ invariant 
mass spectra. Because of the pseudoscalar nature of the $Y$-particle we 
do not expect its formation in $e^+e^-$ collisions.

It is worth noting that in the mass region of interest there
exists one meson, the pseudoscalar-isoscalar $\eta\hsp (1295)$\,, which
cannot be interpreted as a conventional $q\ov{q}$ state~\cite{Kle04}.
It was seen in the reactions $\pi^{-}p\to \eta\pi^{+}\pi^{-}n
(\eta\to \gamma\gamma)$ and 
$\pi^{-}p\to K^{+}K^{-}\pi^0n$ studied in Refs. \cite{Stanton} and 
\cite{E852}, respectively.  
The width of this meson is estimated to $50\div70$ MeV, but
its decay channels are still poorly known. In principle,
$\eta\hsp (1295)$ can be identified with the $Y$-particle. However,
such interpretation would require a more detailed analysis of 
the~$\eta\hsp (1295)$ and $Y$ decay modes.

We believe that $Y$-baryonium can be naturally produced in
ultrarelativistic heavy--ion collisions e.g. at the RHIC in Brookhaven.
It is expected that a nearly baryon--free quark--gluon plasma is formed
at an intermediate stage of such collisions. Then
multi--quark-antiquark clusters should be also formed with a certain
probability. Analysis of experimental data shows~\cite{Bec04,And04}
that ratios of hadron multiplicities observed in central collisions of
nuclei in a broad range of bombarding energies are well reproduced
within a simple thermal model. According to this model the multiplicity 
$N_i$ of hadron species $i$ with mass $m_i$ and spin $J_i$ is
expressed as
\bel{mult}
N_i=\frac{(2J_i+1)\hsp V}{(2\pi)^3}\gamma_s^{n_s}\int {\rm d}^3p
\hsp\left[\exp{\left(\frac{\sqrt{m_i^2+p^2}-\mu_i}{T}\right)}\pm 1
\right]^{-1}\,,
\ee
where $+(-)$ corresponds to fermions (bosons).  It is assumed that all
hadrons are formed in an equilibrated system characterized by volume
$V$ and temperature $T$\,.  In \re{mult} $\mu_i$ stands for the chemical
potential of corresponding hadrons. Following Refs.~\cite{Let99,Bec04}
we take into account possible deviations from chemical equilibrium for
hadrons containing nonzero number ($n_s$) of strange quarks and
antiquarks, by introducing a strangeness suppression factor $\gamma_s$.
On the other hand, approximately the same degree of agreement with
experimental data is achieved in Ref.~\cite{And04} assuming
$\gamma_s=1$\,. Numerical values of the parameters used in
Refs.~\cite{Bec04} (set B) and~\cite{And04} for central Au+Au and Pb+Pb
collisions at different c.m. energies $\sqrt{s}$ are given in Table~I.

\begin{table}[b]
\caption{Fitted parameters of thermal model for central Au+Au
and Pb+Pb collisions at AGS, SPS and RHIC energies.}
\vspace*{3mm}
\begin{ruledtabular}
\begin{tabular}{c|c|c|c|c|c|c}
reaction &$\sqrt{s}$\,(A\hsp GeV)&$T$\,(MeV)~\cite{Bec04} &
$\mu_B$\,(MeV)~\cite{Bec04}&
$\gamma_s$~\cite{Bec04}&$T$\,(MeV)~\cite{And04}&
$\mu_B$\,(GeV)~\cite{And04}\\
\colrule
Au+Au & 4.88 & 119.1 & 578   & 0.763 & 125 & 540\\
Pb+Pb & 8.87 & 145.5 & 375.4 & 0.807 & 148 & 400\\
Pb+Pb & 12.4 & 151.9 & 288.9 & 0.766 &          \\
Pb+Pb & 17.3 & 154.8 & 244.5 & 0.938 & 170 & 255\\
Au+Au & 130  & 165   & 41    & 1.0   & 176 & 41
\end{tabular}
\end{ruledtabular}
\end{table}

Below we estimate multiplicity of $Y$-particles, which might be produced
in central A+A collisions, proceeding from the ratio $N_Y/N_\phi$
predicted by the thermal model. 
This choice is motivated by two reasons. First, $Y$ and
$\phi$ contain the same number of $s\ov{s}$ pairs ($n_s=2$) and,
therefore, $\gamma_s$ factors cancel in the ratio $N_Y/N_\phi$.
Second, their masses are close and as a consequence, the canonical
suppression factors~\cite{And04}, which become important at low
energies, are also approximately cancelled.

In the Boltzman limit, taking into account that \mbox{$\mu_Y=\mu_\phi=0$}
and substituting \mbox{$J_Y=0, J_\phi=1$} into~\re{mult}, one has:
\bel{multr}
\frac{N_Y}{N_\phi}=\frac{1}{3}\left(\frac{m_Y}{m_\phi}\right)^2
\frac{K_2(m_Y/T)}{K_2(m_\phi/T)}\,,
\ee
where $K_2(z)$ is a modified Bessel function of the second order.

\begin{figure*}[htb!]
\vspace*{-9cm}
\hspace*{1cm}\includegraphics[width=13cm]{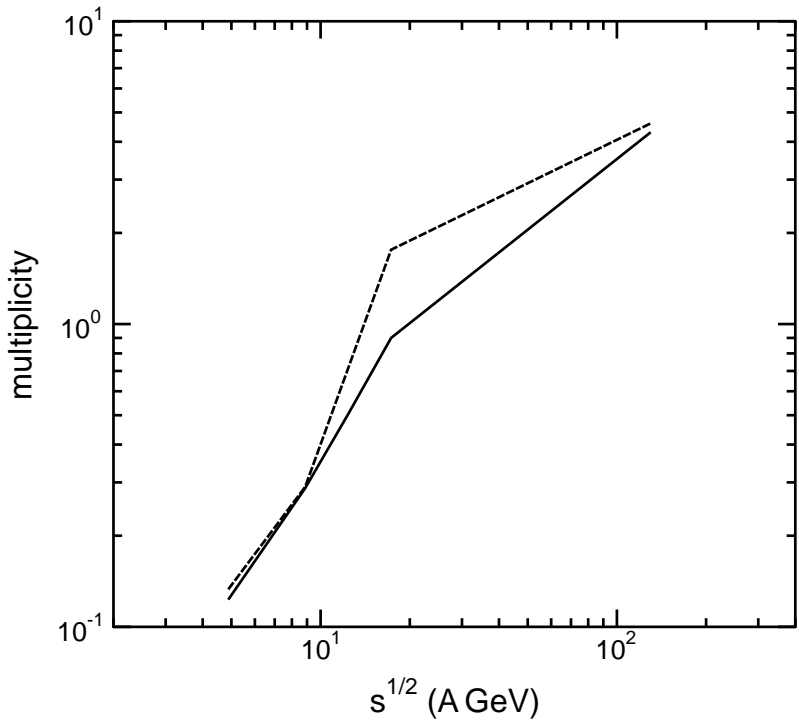}
\caption{Average multiplicities of $Y$-particles
in central Au+Au and Pb+Pb collisions at different c.m. bombarding
energies. The solid and dashed curves correspond to two sets of
parameters suggested, respectively, in Refs.~\cite{Bec04}
and~\cite{And04}.
}
\label{fig1}
\end{figure*}

Using the parameters of Table~I and multiplicities of $\phi$ mesons
given in Refs.~\cite{Bec04,And04,Bac04}, we calculated absolute
multiplicities of $Y$-particles in central Au+Au (AGS, RHIC) and Pb+Pb
(SPS) collisions. The results are shown in Fig.~\ref{fig1}.
The discrepancy between two estimates at $\sqrt{s}=17.3$\, GeV is
explained by two different data sets on $\phi$ multiplicities reported
by the NA49~\cite{Afa00} and NA50~\cite{Ale03} collaborations.
One can see that at energies $\sqrt{s}\gtrsim 10$\,AGeV the predicted
$Y$-\mbox{multiplicities} are rather high, $N_Y\gtrsim 1$\,, which
makes their experimental observation feasible. As discussed above, the
decay modes, \mbox{$Y\to \pi K\ov{K}$}, $Y\to 2\pi\eta$ and $Y\to 2\gamma$ 
could be used for experimental identification of these particles.

Recently in Ref. \cite{Rand03} a similar consideration was used to estimate 
the abundance of $\theta^+$ pentaquarks in nuclear collisions at RHIC. 
About one $\theta^+$ per unit rapidity was predicted per central Au+Au
collision. According to our estimates, due to the lower mass of
$Y$-particles, their yield at midrapidity should be several times larger.

Finally we want to emphasize that the MIT bag model is a
phenomenological model which nevertheless contains many essential
features required for a description of hadrons. Its accuracy for
systems with large number of quarks and antiquarks is unknown. However,
the large attractive color--magnetic interaction suggests that the
$Y$-baryonium state should exist, even if the calculated mass might be 
only approximate.

In conclusion, if a strongly bound baryonium state predicted by the MIT
bag model exists and its decay width is not too large, it can be
produced at the hadronization of a quark--gluon plasma. Using a simple
thermal model we have estimated yields of these particles in
central collisions of relativistic heavy ions. Expected multiplicities
are large enough, $\sim 5$\,per event at RHIC, for their detection.
This observation would not only prove the existence of a new
exotic hadron, sextaquark, but also provide a very strong evidance in
favor of the quark--gluon plasma formation.\\

This work has been partly supported by the GSI,
the DFG Grant \mbox{436 RUS 113/711/0-1} (Germany), the RFBR Grant 03--02--04007 
and the MIS Grant \mbox{NSH--1885.2003.2} (Russia).
I.N.M and D.S. are grateful to the Humboldt Foundation for the
financial support.

\end{document}